%% file: pum_gabidulin.tex
\documentclass[onecolumn]{IEEEtran}

\usepackage[utf8]{inputenc}
\usepackage{graphicx}
\usepackage{amssymb}
\usepackage[cmex10]{amsmath}
\usepackage{color}
\usepackage{theorem}
\usepackage{pifont}
\usepackage{hyperref}
\usepackage{url}
\usepackage{psfrag}
\usepackage{tikz}
\usepackage{bbm}

\usepackage[ruled,vlined,titlenumbered]{algorithm2e}

\input{antoniasdefs.tex}

\IEEEoverridecommandlockouts

\begin{document}

\title{On (Partial) Unit Memory Codes Based on Gabidulin Codes}

\author{\IEEEauthorblockN{Antonia Wachter$^1$, Vladimir Sidorenko$^1$, Martin Bossert$^1$, Victor Zyablov$^2$}\\
\IEEEauthorblockA{$^1$ Institute of Telecommunications and Applied Information Theory, University of Ulm, Ulm, Germany\\
$^2$ Institute for Information Transmission Problems, Russian Academy of Sciences, Moscow, Russia\\}
\texttt{\{antonia.wachter, vladimir.sidorenko, martin.bossert\}@uni-ulm.de, zyablov@iitp.ru}\\
\thanks{This work was supported by the German Research Council "Deutsche Forschungsgemeinschaft" (DFG) under Grant No. Bo867/21-1. V. Sidorenko is on leave from IITP, Russian Academy of Sciences, Moscow, Russia.}}
\maketitle

\begin{abstract}
(Partial) Unit Memory ((P)UM) codes provide a powerful possibility to construct convolutional codes based on block codes in order to achieve a high decoding performance. 
In this contribution, a construction based on Gabidulin codes is considered. This construction requires a modified rank metric, the so--called \emph{sum rank metric}. For the sum rank metric, the \emph{free rank distance}, the \emph{extended row rank distance} and its \emph{slope} are defined analogous to the extended row distance in Hamming metric. 
Upper bounds for the free rank distance and the slope of (P)UM codes in the sum rank metric are derived and an explicit construction of (P)UM codes based on Gabidulin codes is given, achieving the upper bound for the free rank distance. 
\end{abstract}
%
%
\section{Introduction}
A special class of rate $k/n$ convolutional codes --- the so--called \emph{Unit Memory} (UM) codes --- were introduced by Lee in 1976 \cite{Lee_UnitMemory_1976}. UM codes are convolutional codes with memory $m=1$ and overall constraint length $\nu=k$. In \cite{Lauer_PUM_1979}, Lauer extended this idea to \emph{Partial Unit Memory} (PUM) codes where the memory of the convolutional code is also $m=1$, but the overall constraint length is $\nu<k$.

(P)UM codes are constructed based on block codes, e.g. \emph{Reed--Solomon} (RS) \cite{zyablov_sidorenko_pum,Pollara_FiniteStateCodes,Justesen_BDDecodingUM} or BCH codes \cite{DettmarSorger_PUMBCH, DettmarShav_NewUMCodes}. The use of block codes makes an algebraic description of these convolutional codes possible. A convolutional code can be characterized by its \emph{free distance} and the \emph{average linear increase (slope)} of the \emph{extended row distance}. These distance measures determine the error--correcting capability of the convolutional code. In \cite{Lee_UnitMemory_1976}, \cite{Lauer_PUM_1979}, \cite{Pollara_FiniteStateCodes}, \cite{Thommesen_Justesen_BoundsUM} upper bounds for the free (Hamming) distance and the slope of (P)UM codes were derived. There are constructions that achieve the upper bound for the free (Hamming) distance, e.g. \cite{Pollara_FiniteStateCodes}, \cite{Justesen_BDDecodingUM}, \cite{zyablov_sidorenko_pum}. A construction of (P)UM codes based on RS codes that achieves this optimal free distance and also half the optimal slope was given in \cite{zyablov_sidorenko_pum}. 

Rank metric codes recently have attracted attention since they provide an almost optimal solution for error control in random linear network coding \cite{koetter_kschischang}, \cite{silva_rank_metric_approach}. Gabidulin codes are an important class of rank metric codes, they are the rank metric analogues of RS codes and were introduced by Delsarte \cite{Delsarte_1978}, Gabidulin \cite{Gabidulin_TheoryOfCodes_1985} and Roth \cite{Roth_RankCodes_1991}. 

In the context of network coding, dependencies between different blocks transmitted over a network can be created by \emph{convolutional} codes. For multi--shot network coding, where the network is used $N$ times to transmit $N$ blocks, the dependencies between the different shots help to correct more errors than using the classical approach based on rank metric block codes. First approaches for convolutional rank metric codes were given in \cite{BochKudryBossSid_2010_ConvRank}.

In this paper, we consider (P)UM codes based on rank metric block codes. We use the \emph{sum rank metric} for our convolutional codes that is motivated by multi--shot network coding \cite{NobregaBartolomeu_MultishotRankCodes}. For the sum rank metric, we define the \emph{free rank distance} and the \emph{extended row rank distance}. 
We derive upper bounds for the free rank distance and the slope of the extended row rank distance. 
Finally, we give a construction of UM and PUM codes based on Gabidulin codes. We prove that the proposed PUM construction with dual memory $m_H=1$ achieves the upper bound for the free rank distance and half the optimal slope. 

This paper is organized as follows. Section~\ref{sec:def} provides notations and definitions concerning rank metric and convolutional codes. In Section~\ref{sec:pum}, we give the definition of (P)UM codes and show restrictions on the code rate. In Section~\ref{sec:distdefbound}, we define distance measures for convolutional codes based on the sum rank metric and derive upper bounds for the free rank distance and the slope. We give an explicit construction of (P)UM codes based on Gabidulin codes in Section~\ref{sec:constr} and calculate the free rank distance and the slope of the PUM construction. Section~\ref{sec:concl} gives a conclusion.

\section{Definitions and Notations}\label{sec:def}
\subsection{Rank Metric and Gabidulin Codes}
Let $q$ be a power of a prime and let us denote the Frobenius $q$--power for an integer $i$ by $x^{[i]}\defeq x^{q^i}$. Let $\Fq$ denote the finite field of order $q$ and $\F =\Fqs$ its extension field of order $q^s$.

Throughout this paper, let the rows and columns of an $m\times n$--matrix $\mathbf A$ be indexed by $1,\dots, m$ and $1,\dots, n$. 

Given a basis $\mathcal B=\lbrace b_1,b_2, \dots, b_s \rbrace$ of $\F$ over $\Fq$, for any vector $\mathbf a \in \F^n$ there is a one-to-one mapping on an $s \times n$ matrix $\mathbf A$ with entries from $\Fq$. 
The \emph{rank norm} of $\mathbf a \in \F^n$ is defined to be the rank of $\mathbf A$ over $\Fq$:
\begin{equation}\label{eq:ranknorm}
\rank_q(\mathbf a) \defeq \rank (\mathbf A).
\end{equation}

A special kind of matrix that often occurs in the context of Gabidulin codes, will be introduced as follows. 
Let a vector $\mathbf a = \left( \begin{array}{cccc}a_{1} & a_{2} & \dots &\ a_{n} \end{array}\right)$ over $\F$ be given, 
denote the following $m \times n$--matrix by $\mathcal V_m(\mathbf a)$:
\begin{equation}\label{eq:linvand}
 \mathcal V_m (\mathbf a) =\mathcal V_{m} \left( \begin{array}{cccc}a_{1} & a_{2} & \dots &\ a_{n} \end{array}\right)\defeq\left( \begin{array}{cccc}
a_{1} & a_{2} & \dots& a_{n}\\
a_{1}^{[1]} & a_{2}^{[1]} & \dots& a_{n}^{[1]}\\
\vdots&\vdots&\vdots&\vdots\\
a_{1}^{[m-1]} & a_{2}^{[m-1]} & \dots& a_{n}^{[m-1]}\\
\end{array}
\right).
\end{equation}
%
%
%
We call this matrix an $m \times n$ \emph{Frobenius matrix}. If the elements $\lbrace a_1,\dots,a_n \rbrace \in \F$ are linearly independent over $\Fq$ (i.e., if $\rank_q(\mathbf a)=s)$), then $\mathcal V_m (\mathbf a)$ always has full rank $\min \lbrace m,n\rbrace$ over $\F$ and \emph{any} $\min \lbrace m,n\rbrace$ columns are linearly independent \cite[Lemma~3.5.1]{Lidl-Niederreiter:FF1996}.



Gabidulin codes are a special class of rank metric codes. They are the rank metric analogues of RS codes and hence, there exist efficient decoding strategies similar to the ones for RS codes. A Gabidulin code is defined by its parity--check matrix as follows. 

\begin{definition}[Gabidulin Code \cite{Gabidulin_TheoryOfCodes_1985}]
A linear $(n,k)$ Gabidulin code $\mathcal{G}$ over $\F$ for $n \leq s$ is defined by its $(n-k) \times n$ parity check matrix $\mathbf H$:
\begin{equation*}
\mathbf H = \mathcal V_{n-k} (\mathbf h) = \mathcal V_{n-k} \left( \begin{array}{cccc}h_{1} & h_{2} & \dots &\ h_{n} \end{array}\right),
\end{equation*}
where the elements $\lbrace h_1, \dots, h_n \rbrace \in \F$ are linearly independent over $\Fq$. 
\end{definition}

The \textit{minimum rank distance} $d_{\rk}$ of a linear block code $\mathcal C$ over $\F$ is defined by:
\begin{equation*}
d_{\rk} \defeq \min \lbrace \rank_q(\mathbf c) \; | \; \mathbf c \in \mathcal C, \mathbf c \neq \mathbf 0 \rbrace.
\end{equation*}
Gabidulin codes are \emph{Maximum Rank Distance} (MRD) codes, i.e., they fulfill the rank metric equivalent of the Singleton bound with equality and achieve $d_{\rk}=n-k+1$ \cite{Gabidulin_TheoryOfCodes_1985}.

\subsection{Convolutional Codes}
\begin{definition}[Convolutional Code]
A rate $k/n$ convolutional code $\mathcal C$ over $\F$ with memory $m$ (see Definition~\ref{def:memory}) is defined by its $k \times n$ generator matrix in polynomial form:
\begin{equation*}
\mathbf G (D) = \left( \begin{array}{cccc}
g_{11}(D) & g_{12}(D)&\dots&g_{1n}(D)\\
\vdots&\vdots&\ddots&\vdots\\
g_{k1}(D) & g_{k2}(D)&\dots&g_{kn}(D)\\
\end{array}\right),
\end{equation*}
where
\begin{equation*}
g_{ij}(D) = g_{ij}^{(0)} + g_{ij}^{(1)}D + g_{ij}^{(2)}D^2 +\dots+ g_{ij}^{(m)}D^m, \qquad g_{ij}^{(\ell)} \in \F, \ \ell=0,\dots,m,
\end{equation*}
for $i=1,\dots,k$ and $j=1,\dots,n$. 
\end{definition}

There exist different definitions of memory and constraint length in the literature; we denote the \emph{constraint lengths} $\nu_i$, the \emph{memory} $m$ and the \emph{overall constraint length} $\nu$ of the convolutional code $\mathcal C$ according to \cite{Johannesson_Fund_Conv_Codes} as follows.
\begin{definition}[Memory and Constraint Length]\label{def:memory}
The constraint length for the $i$--th input of a polynomial generator matrix $\mathbf G(D)$ is
\begin{equation*}
\nu_i \defeq \max_{\substack{1 \leq j \leq n}} \lbrace \deg g_{ij}(D) \rbrace.
\end{equation*}
The memory is the maximum constraint length:
\begin{equation*}
m \defeq \max_{\substack{1\leq i \leq k}} \lbrace \nu_i \rbrace.
\end{equation*}
The overall constraint length $\nu$ is defined as the sum of the constraint lengths $\nu_i$:
\begin{equation*}
\nu \defeq \sum \limits_{i=1}^{k} \nu_i.
\end{equation*}
\end{definition}

A codeword $\mathbf c(D)=c_0 +c_1 D + c_2 D^2 +\dots$ of the convolutional code $\mathcal C$ is generated by
\begin{equation*}
\mathbf c(D)=\mathbf u(D) \cdot \mathbf G(D),
\end{equation*}
where $\mathbf u(D)=u_0 +u_1 D + u_2 D^2 +\dots$ is the information word.

A polynomial $(n-k)\times n$ parity--check matrix $\mathbf H(D)$ of $\mathcal C$ is defined such that for every codeword $\mathbf c(D) \in \mathcal C$, 
\begin{equation*}
\mathbf c(D) \cdot \mathbf H^T(D) = 0,
\end{equation*}
where
\begin{equation*}
\mathbf H (D) = \left( \begin{array}{cccc}
h_{11}(D) & h_{12}(D)&\dots&h_{1n}(D)\\
\vdots&\vdots&\ddots&\vdots\\
h_{(n-k)1}(D) & h_{(n-k)2}(D)&\dots&h_{(n-k)n}(D)\\
\end{array}\right),
\end{equation*}
with
\begin{equation*}
h_{ij}(D) = h_{ij}^{(0)} + h_{ij}^{(1)}D + h_{ij}^{(2)}D^2 +\dots+ h_{ij}^{(m_H)}D^{m_H}, \qquad h_{ij}^{(\ell)} \in \F \ \ell=0,\dots,m_H.
\end{equation*}

We can also represent $\mathbf G(D)$ and $\mathbf H(D)$ as semi--infinite matrices over $\F$:
\begin{equation}\label{eq:def_semiinf_H}
\mathbf G = \left( \begin{array}{cccccc}
\mathbf G_0 &\mathbf G_1& \dots& \mathbf G_m && \\
&\mathbf G_0 &\mathbf G_1& \dots&\mathbf  G_m& \\
&&\ddots&\ddots&\ddots&\ddots\\
             \end{array}\right), \qquad
\mathbf H = \left( \begin{array}{ccc}
\mathbf H_0 && \\
\mathbf H_1 &\mathbf H_0& \\
\mathbf H_2 &\mathbf H_1&\\
\vdots&\vdots&\ddots\\
\mathbf H_{m_H} & \mathbf H_{m_H-1}&\ddots\\
&\mathbf H_{m_H} &\ddots\\
&&\ddots\\
             \end{array}\right),
\end{equation}
where $\mathbf G_i$, $i=0,\dots,m$ are $k \times n$--matrices and $\mathbf H_i$, $i=0,\dots,m_H$ are $(n-k)\times n$ matrices. Note that in general the number of submatrices in $\mathbf H$ and $\mathbf G$, is not equal, i.e., $m_H \neq m$. Let us denote $m_H$ as the \emph{dual memory} of $\mathcal C$. If both $\mathbf G$ and $\mathbf H$ are in minimal basic encoding form, the overall number of memory elements, i.e., the overall constraint length $\nu$ is the same in both representations \cite{Forney_ConvolutionalCodesAlg}.

$\mathbf c(D)$ and $\mathbf u(D)$ can be represented as causal infinite sequences:
\begin{equation*}
\mathbf c = \left( \begin{array}{cccc}	\mathbf c_0& \mathbf c_1& \mathbf c_2 &\dots \end{array}\right), \qquad \mathbf u= \left( \begin{array}{cccc}	\mathbf u_0& \mathbf u_1& \mathbf u_2 &\dots \end{array}\right),
\end{equation*}
where $\mathbf c_j \in \F^n$ and $\mathbf u_j \in \F^k$ for all $j$. Then, $\mathbf c = \mathbf u \cdot \mathbf G$ and $\mathbf c \cdot \mathbf H^T = \mathbf 0$.

\section{(Partial) Unit Memory Codes}\label{sec:pum}
(P)UM codes are a special class of convolutional codes with memory $m=1$ \cite{Lee_UnitMemory_1976}, \cite{Lauer_PUM_1979}, i.e., the semi--infinite generator matrix $\mathbf G$ is given by:\\
\begin{equation}\label{eq:def_Gm1}
\mathbf G = \left( \begin{array}{ccccc}
\mathbf G_0 &\mathbf G_1&& \\
&\mathbf G_0 &\mathbf G_1& \\
&&\ddots&\ddots\\
             \end{array}\right),
\end{equation}
where $\mathbf G_0$ and $\mathbf G_1$ are $k\times n$ matrices. For an $(n,k)$ UM code, both matrices have full rank. 
For an $(n,k\;|\;k_1)$ PUM code, $\rank(\mathbf G_0)=k$ and $\rank(\mathbf G_1)=k_1 < k$:
\begin{equation*}
\mathbf G_0 = \left( \begin{array}{c} \mathbf G_{00}\\ \mathbf G_{01}\end{array}\right), \qquad \mathbf G_1 = \left( \begin{array}{c} \mathbf G_{10}\\ \mathbf 0\end{array}\right),
\end{equation*}
where $\mathbf G_{00}$ and $\mathbf G_{10}$ are $k_1 \times n$ matrices and $\mathbf G_{01}$ is $(k-k_1)\times n$--matrix. 

Hence, for both cases, we have the following encoding rule:
\begin{equation*}
\mathbf c_j = \mathbf u_j \cdot \mathbf G_0 + \mathbf u_{j-1} \cdot \mathbf G_1.
\end{equation*}

Note that the overall constraint length for UM codes is $\nu = k$ and for PUM codes $\nu = k_1$. 

As will be shown in the following, there are restrictions on the code rate of (P)UM codes when a certain number of full--rank submatrices $\mathbf H_i$ \eqref{eq:def_semiinf_H} should exist. This full--rank condition, $\rank(\mathbf H_i)=n-k$ for all $i=0,\dots,m_H$ is used in our construction.

\begin{lemma}[Rate Restriction for UM Codes]
An $(n,k)$ UM code with overall constraint length $\nu=k$ has rate 
\begin{equation*}
R=\frac{(n-k)\cdot m_H}{(n-k)\cdot(m_H+1)} 
\end{equation*}
if the parity--check matrix $\mathbf H$ in minimal basic encoding form consists of $m_H+1$ full--rank submatrices $\mathbf H_i$ \eqref{eq:def_semiinf_H} for $m_H \geq 1$.
\end{lemma}
\begin{IEEEproof}
The overall constraint length $\nu$ has to be the same in representation of a code by $\mathbf G$ and $\mathbf H$ if both are in minimal basic encoding form \cite{Forney_ConvolutionalCodesAlg}. Since $\rank(\mathbf H_i)=n-k$ for all $i=0,\dots,m_H$, we have $\nu=m_H\cdot(n-k)$. On the other hand, the UM code is defined by $\mathbf G$ such that $\nu = k$, hence,
\begin{equation*}
k=m_H\cdot(n-k)  \quad \Longleftrightarrow \quad  R = \frac{k}{n} = \frac{(n-k) \cdot m_H}{(n-k)\cdot (m_H+1)}.
\end{equation*}
\end{IEEEproof}

\begin{lemma}[Rate Restriction for PUM Codes]
Let an $(n,k\;|\;k_1)$ PUM code with $\nu=k_1<k$ be given. Its rate is 
\begin{equation*}
R =\frac{k}{n}> \frac{m_H}{m_H+1},
\end{equation*}
if the parity--check matrix $\mathbf H$ in minimal basic encoding form consists of $m_H+1$ submatrices $\mathbf H_i$ \eqref{eq:def_semiinf_H} for $m_H \geq 1$.
\end{lemma}
\begin{IEEEproof}
The overall constraint length $\nu$ is the same in the representation of a code by $\mathbf G$ and $\mathbf H$ in minimal basic encoding form \cite{Forney_ConvolutionalCodesAlg}. Since $\rank(\mathbf H_i)=n-k$ for all $i$, we have $\nu=m_H\cdot(n-k)$. The PUM code is defined such that $\nu=k_1 <k$, hence,
\begin{equation*}
m_H\cdot(n-k) < k  \quad \Longleftrightarrow \quad  R = \frac{k}{n} > \frac{m_H}{m_H+1}.
\end{equation*}
\end{IEEEproof}

If we use the \emph{parity--check} matrix to construct (P)UM codes, the following theorem guarantees that there is always a corresponding generator matrix that defines a (P)UM code.

\begin{theorem}\label{theo:HdefsPUM}
 For each semi--infinite parity--check matrix $\mathbf H$, where the $(m_H+1)$ submatrices $\mathbf H_i$ \eqref{eq:def_semiinf_H} are $(n-k) \times n$ matrices of $\rank(\mathbf H_i)=n-k$, $\forall i$, and $R\geq m_H/(m_H+1)$, there exists a generator matrix $\mathbf G$ such that $\mathbf G \cdot \mathbf H^T=\mathbf 0$ and $\mathbf G$ defines a (P)UM code \eqref{eq:def_Gm1} with $k_1 \geq (k+1)/2$.
\end{theorem}
\begin{IEEEproof}
For (P)UM codes, the generator matrix $\mathbf G$ consists of two submatrices $\mathbf G_0$ and $\mathbf G_1$. In order to show that there is always such a $\mathbf G$, we show that there are more unknown entries in these two matrices than equations that have to be fulfilled. 
The condition $\mathbf G \cdot \mathbf H^T=\mathbf 0$ corresponds to the following equations:
\begin{align*}
 &\mathbf G_0 \cdot\mathbf H_{0}^T=\mathbf 0, \\
&\mathbf G_0 \cdot\mathbf H_{i}^T+\mathbf G_1\cdot \mathbf H_{i-1}^T=\mathbf 0, \qquad \forall \ i=1,\dots,m_H, \\
 &\mathbf G_1 \cdot\mathbf H_{m_H}^T=\mathbf 0.
\end{align*}
Overall, there are at most $(m_H+1)k(n-k)+k_1(n-k)$ linearly independent equations with $k_1 \leq k$. 
Additionally, we have to guarantee that $\rank(\mathbf G_0)=k$. This is always fulfilled if an arbitrary $k\times k$--submatrix of $\mathbf G_0$ is a lower or upper triangular matrix. Thus, we need $\sum_{i=1}^{k-1} i = k(k-1)/2$ zero entries at certain positions of $\mathbf G_0$ and hence we have additional $k(k-1)/2$ equations that have to be fulfilled. 
There are $(k+k_1)n$ unknown entries of $\mathbf G$. Hence, the number of entries $g_{free}$ that can be chosen freely is given by:
\begin{equation*}
g_{free}\geq (k+k_1)n-(m_H+1)k(n-k)-k_1(n-k)-(k(k-1))/2
=k\big((m_H+1)k+k_1-m_H n-(k-1)/2\big).
\end{equation*}
Since $R\geq m_H/(m_H+1)$,
\begin{equation*}
 g_{free}\geq k\Big((m_H+1)\frac{m_H \cdot n}{m_H+1}+k_1-m_H n-(k-1)/2)\Big)=k \left(k_1-(k-1)/2\right).
\end{equation*}
Hence, $g_{free} \geq 0$ if $k_1 \geq (k+1)/2$ and the statement follows.
\end{IEEEproof}

\section{Distance Measures for Convolutional Codes Based on Rank Metric}\label{sec:distdefbound}
\subsection{Distance Definitions}\label{subsec:distdef}
In this section, we define distance measures for convolutional codes based on a special rank metric. This special metric is the \emph{sum rank metric} that is used in \cite{NobregaBartolomeu_MultishotRankCodes} for multi--shot uses of a network. In \cite{NobregaBartolomeu_MultishotRankCodes}, they denote the sum rank metric by \emph{extended rank metric} and introduce furthermore the \emph{extended version of the lifting construction}. They show that the sum rank distance and the subspace distance of the extended version of the lifting construction have the same relation as the \emph{rank distance} and the subspace distance of the \emph{lifting construction} \cite{silva_rank_metric_approach}. Hence, a code constructed by the extended lifting construction keeps the distance properties of the underlying code based on the sum rank metric and the use of the sum rank metric for multi--shot network coding can be seen as the analogue to using the rank metric for single--shot network coding. This property motivates the use of the sum rank metric. First, the sum rank weight is defined as follows.

\begin{definition}[Sum Rank Weight]
Let a vector $\mathbf v \in \F^{\ell n}$ be given and let it be decomposed into subvectors: 
\begin{equation*}
\mathbf v = \left( \begin{array}{cccc}\mathbf v_0 & \mathbf v_1& \dots& \mathbf v_{\ell-1} \end{array}\right),
\end{equation*}
with $\mathbf v_i \in \F^n$ for all $i$. We define the \emph{sum rank weight} $\wt_{rk}(\mathbf v)$ as the \emph{sum} of the rank norms \eqref{eq:ranknorm} of the subvectors:
\begin{equation}\label{eq:defsumrankweight}
\wt_{rk}(\mathbf v) \defeq \sum \limits_{i=0}^{\ell-1} \rank_q (\mathbf v_i),
\end{equation}
for $ 0 \leq \ell \leq \infty$.
\end{definition}

Hence, we define the \emph{sum rank distance} between two sequences $\mathbf v^{(1)},\mathbf v^{(2)}$ of length $\ell n$ by
\begin{equation}\label{eq:defsumrankdist}
d(\mathbf v^{(1)},\mathbf v^{(2)}) \defeq \sum \limits_{i=0}^{\ell-1}\wt_{rk}(\mathbf v^{(1)}_i-\mathbf v^{(2)}_i)=\sum \limits_{i=0}^{\ell-1}\rank_q(\mathbf v^{(1)}_i-\mathbf v^{(2)}_i)=\sum \limits_{i=0}^{\ell-1} d_{\rk}(\mathbf v^{(1)}_i,\mathbf v^{(2)}_i).
\end{equation}

An important measure for convolutional codes is the free distance, and consequently we define the \emph{free rank distance} $d_{free}$ as follows.
\begin{definition}[Free Rank Distance]
The minimum sum rank distance \eqref{eq:defsumrankdist} between any two nonzero codewords $\mathbf c^{(1)},\mathbf c^{(2)}$ from a convolutional code $\mathcal C$ is called the \emph{free rank distance} $d_{free}$:
\begin{equation}\label{eq:deffreedist}
d_{free}\defeq
\min_{\substack{\mathbf c^{(1)} \neq \mathbf c^{(2)}}} \left\lbrace d(\mathbf c^{(1)},\mathbf c^{(2)})\right \rbrace
=\min_{\substack{\mathbf c^{(1)} \neq \mathbf c^{(2)}}} \left\lbrace \sum \limits_{i=0}^{\infty} d_{\rk}(\mathbf c^{(1)}_i,\mathbf c^{(2)}_i)\right \rbrace.
\end{equation}
\end{definition}
Note that this definition differs from the definition in \cite{BochKudryBossSid_2010_ConvRank} which is also called \emph{free rank distance}.

Generally, the error--correcting capability of convolutional codes is determined by \emph{active distances}. In the following, we define the \emph{extended row rank distance} which is an important active distance in the sum rank metric. 

Let $\mathcal C^r(\ell)$ denote the set of all codewords $\mathbf c^{(\ell)}$ corresponding to paths in the code trellis which diverge from the zero state at depth $j$ and return to the zero state for the \emph{first} time after $\ell+1$ branches at depth $j+\ell+1$. Without loss of generality, we assume $j=0$ as we only consider time--invariant convolutional codes. (P)UM codes can be represented by a trellis where the state at depth $i$ is the information subvector $\mathbf u_{i-1}$, i.e., the zero state can always be reached in one step with the input $\mathbf u_i=\mathbf 0$ \cite{Thommesen_Justesen_BoundsUM}. Therefore, a path in the trellis is not merged with the zero part if and only if each information subvector $\mathbf u_i$ is nonzero (and hence also each $\mathbf c_i$). Thus, in order to reach the zero state for the first time in depth $\ell+1$, for all codewords $\mathbf c^{(\ell)}$ in $\mathcal C^r(\ell)$, we have $\mathbf c_{\ell}=\mathbf 0$ and:
\begin{equation}\label{eq:defcl}
\mathbf c^{(\ell)}= \left( \begin{array}{ccccccccc}\dots & \mathbf 0& \mathbf c_{0}& \mathbf c_{1} &\mathbf c_{2} & \dots &\mathbf c _{\ell-1}&\mathbf 0&\dots \end{array}\right),
\end{equation}
with $\mathbf c_i \neq 0 $ for $i=0,\dots,\ell-1$ and the path corresponding to $\left( \begin{array}{ccccccc} \mathbf 0& \mathbf c_{0}& \mathbf c_{1} &\mathbf c_{2} & \dots &\mathbf c _{\ell-1}&\mathbf 0 \end{array}\right)$ has length $\ell+1$.

The extended row rank distance of order $\ell$ is defined as the minimum sum rank weight of all codewords in $\mathcal C^r(\ell)$.
\begin{definition}[Extended Row Rank Distance]
The extended row rank distance of order $\ell=1,2,\dots$ is defined as
\begin{equation}\label{eq:defextrowrank}
d^r_{\ell} \defeq \min_{\substack{\mathbf c \in \mathcal C^r(\ell)}} \big\lbrace \wt_r(\mathbf c) \big\rbrace.
\end{equation}
\end{definition}

The minimum of the $\ell$th order extended row rank distances gives the free rank distance \eqref{eq:deffreedist}:
\begin{equation*}
d_{free} = \big\lbrace\min_{\substack{\ell}} d^r_{\ell}\big\rbrace.
\end{equation*}
As for Hamming metric, the extended row rank distance $d^r_{\ell}$ can be lower bounded by a linear function $d^r_{\ell} \geq \max \lbrace \alpha \ell +\beta, d_{free} \rbrace$ where $\beta \leq d_{free}$ and $\alpha$ denotes the \emph{slope} (the average linear increase). The slope is an important parameter for determining the error--correcting capability and is defined as follows.
\begin{definition}[Slope]
The average linear increase of the extended row rank distance (slope) is defined as
\begin{equation}\label{eq:defslope}
\alpha \defeq \lim_{\substack{\ell \rightarrow \infty}} \left \lbrace \frac{d^r_{\ell}}{\ell} \right \rbrace.
\end{equation}
\end{definition}

\subsection{Upper Bounds for Distances of (P)UM Codes Based on the Sum Rank Metric}\label{sec:upperbound}
In this section, we derive upper bounds for the free rank distance \eqref{eq:deffreedist} and the slope \eqref{eq:defslope} for UM and PUM codes based on the sum rank metric \eqref{eq:defsumrankweight}, \eqref{eq:defsumrankdist}. The derivation of the bounds uses the well--known bounds for (P)UM codes based on Hamming metric block codes \cite{Lee_UnitMemory_1976}, \cite{Lauer_PUM_1979}, \cite{Pollara_FiniteStateCodes}. 

Assume, the codewords of a convolutional code are considered in \emph{Hamming} metric. Let us then denote the $\ell$th order active row \emph{Hamming} distance by $d^{H,r}_{\ell}$ and the free \emph{Hamming} distance of this convolutional code by $d_{free}^H$. The following theorem provides connections between the free \emph{rank} distance $d_{free}$ and the free \emph{Hamming} distance $d_{free}^H$ and between the extended row \emph{rank} distances $d^r_{\ell}$ and the extended row \emph{Hamming} distances $d^{H,r}_{\ell}$.
\begin{theorem}[Connection between Distances in Hamming and Sum Rank Metric]
For the free rank/Hamming distance and the extended row rank/Hamming distance, the following holds:
\begin{align*}
d_{free} &\leq d_{free}^H,\\
d^r_{\ell} &\leq d^{H,r}_{\ell}, \quad \ell=1,2,\dots.
\end{align*}
\end{theorem}
\begin{IEEEproof}
Due to \cite[Lemma~1]{Gabidulin_TheoryOfCodes_1985}, the following holds for the rank norm $\rank_q(\mathbf v_i)$ and the Hamming norm $\wt_H(\mathbf v_i)$ of a vector $\mathbf v_i$:
\begin{equation*}
\rank_q(\mathbf v_i) \leq \wt_H(\mathbf v_i).
\end{equation*}
and hence also
\begin{equation*}
\sum\limits_{i=0}^{\ell-1}\rank_q(\mathbf v_i) \leq \wt_H(\mathbf v_0 \dots \mathbf v_{\ell-1}).
\end{equation*}
and the statement follows with \eqref{eq:defsumrankweight}, \eqref{eq:deffreedist}, \eqref{eq:defextrowrank}.
\end{IEEEproof}
Consequently, the upper bounds for the free distance and the slope of (P)UM codes based on Hamming metric \cite{Thommesen_Justesen_BoundsUM}, \cite{Pollara_FiniteStateCodes} also hold for (P)UM codes based on the sum rank metric.

\begin{corollary}[Upper Bounds]\label{cor:upperbounds}
For an $(n,k)$ UM code, where $\nu=k$, the free rank distance is upper bounded by:
\begin{equation}\label{eq:umupperbound}
d_{free} \leq 2n-k+1.
\end{equation}
For an $(n,k \;| \; k_1)$ PUM code, where $\nu=k_1 < k$, the free rank distance is upper bounded by:
\begin{equation}\label{eq:pumupperbound}
d_{free} \leq n-k+\nu+1.
\end{equation}
For both UM and PUM codes, the average linear increase (slope) is upper bounded by:
\begin{equation}\label{eq:slopeupperbound}
\alpha \leq n-k.
\end{equation}
\end{corollary}

\section{Construction of (P)UM Codes Based on Gabidulin Codes}\label{sec:constr}

\subsection{Construction}

In this section, we construct (P)UM codes based on the parity--check matrices of Gabidulin codes. In Definition~\ref{def:pumgab}, we give some properties that must be fulfilled for a (P)UM code based on Gabidulin codes and afterwards we show an explicit construction using normal bases. 
\begin{definition}[(P)UM Code Based on Gabidulin Codes]\label{def:pumgab}
For some $m_H \geq 1$, let a rate $R=k/n= c\cdot m_H/(c\cdot(m_H+1))$ UM code or a rate $R=k/n>m_H/(m_H+1)$ PUM code over $\F$ be defined by its semi--infinite parity--check matrix $\mathbf H$ \eqref{eq:def_semiinf_H}. 
Let each submatrix $\mathbf H_i$, $i=0,\dots,m_H$ be the parity--check matrix of an $(n,k)$ Gabidulin code $\mathcal G^{(i)}$:
\begin{equation*}
\mathbf H_i=  \mathcal V_{n-k} (\mathbf h^{(i)}) = \mathcal V_{n-k}  \big( \begin{array}{cccc} h_1^{(i)}& h_2^{(i)} & \dots & h_n^{(i)} \end{array}\big),\quad \forall\quad i=0,\dots,m_H.
\end{equation*}
Additionally, let
\begin{align}
\mathbf H^{(c)}& \defeq\left( \begin{array}{c} \mathbf H_0\\ \mathbf H_1 \\ \dots \\ \mathbf H_{m_H}\end{array}\right) \quad \text{define} \quad \mathcal G^{(c)},\label{eq:defHc}\\
\mathbf H^{(r(i))}& \defeq\left( \begin{array}{cccc} \mathbf H_i & \mathbf H_{i-1} & \dots & \mathbf H_0\end{array}\right) \quad \text{define} \quad \mathcal G^{(r_i)}, \quad \forall\quad i=1,\dots,m_H,\label{eq:defHr}
\end{align}
where $\mathcal G^{(c)}$ is an $(n^{(c)},k^{(c)})$ Gabidulin code and $\mathcal G^{(r_i)}$ is an $(n^{(r_i)},k^{(r_i)})$ Gabidulin code with 
\begin{align*}
&n^{(c)}=n, \quad k^{(c)}=n-(m_H+1)(n-k),\\
&n^{(r_i)}=(i+1)n,\quad k^{(r_i)}=i\cdot n+k,\quad i=1,\dots,m_H.
\end{align*}
\end{definition}
Hence, not only each submatrix has to define a Gabidulin code, but also the rows and columns of submatrices of $\mathbf H$. 

Note that Theorem~\ref{theo:HdefsPUM} guarantees that for the parity--check matrix from Definition~\ref{def:pumgab} there is always a generator matrix $\mathbf G$ that defines a (P)UM code.

Now, we give an explicit construction that fulfills the requirements of Definition~\ref{def:pumgab}. To ensure that \eqref{eq:defHc} is fulfilled and $\mathbf H^{(c)}$ defines a Gabidulin code, $\mathbf H_1$ has to be the continuation of $\mathbf H_0$, i.e., $\mathbf h^{(1)}=\big((h_1^{(0)})^{[n-k]},(h_2^{(0)})^{[n-k]},\dots, (h_n^{(0)})^{[n-k]}\big)$. In addition, $\mathbf h^{(2)}$ has to be the continuation of $\mathbf h^{(1)}$ and so on. Hence, 
\begin{equation}\label{eq:defHch0}
 \mathbf H^{(c)} = \mathcal V_{(m_H+1)(n-k)} \big(\mathbf h^{(0)}\big)=\mathcal V_{(m_H+1)(n-k)}\big( \begin{array}{cccc}  h_1^{(0)} & h_2^{(0)} & \dots & h_n^{(0)} \end{array}\big).
\end{equation}

In order to fulfill \eqref{eq:defHr}, we have to ensure that all elements from $\F$ in the set 
\begin{equation}\label{eq:setH}
\begin{split}
 \mathcal H &\defeq \lbrace h_1^{(0)},\dots, h_n^{(0)},h_1^{(1)},\dots, h_n^{(1)},\dots,h_1^{(m_H)},\dots, h_n^{(m_H)}\rbrace\\
&=\lbrace h_1^{(0)},\dots, h_n^{(0)},(h_1^{(0)})^{[n-k]},\dots, (h_n^{(0)})^{[n-k]},\dots,(h_1^{(0)})^{[m_H(n-k)]},\dots, (h_n^{(0)})^{[m_H(n-k)]}\rbrace
\end{split}
\end{equation}
with $|\mathcal H|=(m_H+1)\cdot n$ are linearly independent over $\Fq$.

To obtain an explicit construction of such a (P)UM code, we can use a \emph{normal basis}. 
A basis $\mathcal B = \lbrace b_0, b_1, \dots, b_{s-1}\rbrace$ of $\F$ over $\Fq$ is a normal basis if $b_i= b^{[i]}$ for all $i$ and $b\in \F$ is called a \textit{normal element}. There is a normal basis for any finite extension field $\F$ \cite{Lidl-Niederreiter:FF1996}. For our construction, we use a normal element $b$ to define
\begin{equation}\label{eq:defh0normal}
 \mathbf h^{(0)}=\big( \begin{array}{ccc|ccc|ccc|c} b^{[0]} & \dots & b^{[n-k-1]} & b^{[(m_H+1)(n-k)]}&\dots& b^{[(m_H+2)(n-k)-1]}& b^{[2(m_H+1)(n-k)]}&\dots& b^{[2(m_H+2)(n-k)-1]}&\dots \end{array}\big).
\end{equation}
This $\mathbf h^{(0)}$ is used to define $\mathbf H^{(c)}$ \eqref{eq:defHch0} and hence also $\mathbf H$ is defined \eqref{eq:defHc}, \eqref{eq:def_semiinf_H}. 

To make sure that also \eqref{eq:defHr}, \eqref{eq:setH} are fulfilled, we require a certain minimal field size. 
If $(n-k)$ divides $n$, then $\mathbf h^{(0)}$ can be divided into subvectors, each of length $(n-k)$ \eqref{eq:defh0normal} and the field size has to fulfill $s \geq (m_H+1)\cdot n$ to ensure that all elements in $\mathcal H$ are linearly independent \eqref{eq:setH} and hence that \eqref{eq:defHr} is fulfilled. 
If $(n-k)$ does not divide $n$, the last subvector in $\mathbf h^{(0)}$ is shorter than $n-k$. 
Equations \eqref{eq:setH} and \eqref{eq:defHr} can be guaranteed in general if 
\begin{equation}\label{eq:restrFqs}
s \geq (m_H+1)\cdot \left\lceil\frac{n}{n-k}\right\rceil \cdot (n-k).
\end{equation}
This implies the restriction for $(n-k)|n$ as a special case.

The following lemma shows that the parity--check matrix constructed in such a way is in minimal basic form.
\begin{lemma}
 Let a (P)UM code based on Gabidulin codes be defined by its parity--check matrix $\mathbf H$ as in Definition~\ref{def:pumgab}. Then, $\mathbf H(D)$ is in minimal basic form. 
\end{lemma}
\begin{IEEEproof}
 First, we show that $\mathbf H(D)$ is in \emph{basic} form. According to \cite[Definition 4]{Forney_ConvolutionalCodesAlg}, $\mathbf H(D)$ is basic if it is polynomial and if there exists a polynomial right inverse $\mathbf H^{-1}(D)$, such that $\mathbf H(D) \cdot \mathbf H^{-1}(D)=\mathbf I_{(n-k)\times(n-k)}$, where $\mathbf I_{(n-k)\times(n-k)}$ denotes the $(n-k)\times(n-k)$--identity matrix. By definition, $\mathbf H(D)$ is polynomial. A polynomial right inverse exists if and only if $\mathbf H(D)$ is non--catastrophic 
 and hence if the slope $\alpha >0$ \cite[Theorem~A.4]{Dettmar_PUM_PhD_1994}. We will calculate the slope in Theorem~\ref{theo:dfreeslope}, which gives us $\alpha>0$. Hence, $\mathbf H(D)$ is in basic form.

Second, we show that $\mathbf H(D)$ is in \emph{minimal} basic form. According to \cite[Definition 5]{Forney_ConvolutionalCodesAlg}, a basic $(n-k)\times n$ matrix $\mathbf H(D)$ is minimal if its overall constraint length $\nu$ in the obvious realization is equal to the maximum degree $\mu$ of its $(n-k)\times(n-k)$--subdeterminants. We have $\mathbf H(D)=\mathbf H_0 + \mathbf H_1 \cdot D + \dots + \mathbf H_{m_H} \cdot D^{m_H}$ and hence $\deg ( h_{ij}(D))=m_H$ for all $ 1\leq i,j \leq n-k$. Thus, for each $(n-k)\times (n-k)$--subdeterminant starting in column $\ell=1,\dots,k+1$, we know that $\deg [\det(\mathbf H^{(\ell)}(D)) ] \leq m_H(n-k)$. The coefficient of $D^{m_H(n-k)}$ of $\det(\mathbf H^{(\ell)}(D))$ is exactly $\det(\mathbf H^{(\ell)}_{m_H})$, where $\mathbf H^{(\ell)}_{m_H}$ is the $(n-k)\times(n-k)$--submatrix of $\mathbf H_{m_H}$ starting in column $\ell$. Since $\mathbf H^{(\ell)}_{m_H}$ is an $(n-k)\times(n-k)$ Frobenius matrix where the elements in the first row are linearly independent, $\det(\mathbf H^{(\ell)}_{m_H}) \neq 0$ \cite{Lidl-Niederreiter:FF1996} and $\deg [ \det(\mathbf H^{(\ell)}(D)) ] = m_H(n-k)$, $\forall \ell=1,\dots,k+1$. This is equal to the constraint length in obvious realization $\nu=m_H(n-k)$ and hence, $\mathbf H(D)$ is in  minimal basic form.
\end{IEEEproof}

Our construction is demonstrated in an example with $m_H=1$ in the following.

\begin{example}\label{ex:defh}
Let us construct an $(6,4 \;|\; 2)$ PUM code with $m_H=1$. Hence, $s \geq 12$ \eqref{eq:restrFqs} and we define the code e.g. over $\mathbb{F}_{2^{12}}$. In this field, there exists a \emph{normal element} $a$ such that $ a^{[0]},a^{[1]},\dots,a^{[11]} $ are all linearly independent over $\mathbb F_2$ (i.e., this is a \emph{normal basis}).

In order to guarantee also the continuation with $n-k=2$, we choose $\mathbf h^{(0)}$ as in \eqref{eq:defh0normal} and $\mathbf h^{(1)}$ is defined by \eqref{eq:defHc}, \eqref{eq:defHch0}:
\begin{align*}
 \mathbf h^{(0)} &= \left( \begin{array}{cccccc} a^{[0]} & a^{[1]} & a^{[4]} & a^{[5]} &  a^{[8]} & a^{[9]}\end{array}\right),\\
\mathbf h^{(1)} &= \left( \begin{array}{cccccc} a^{[2]} & a^{[3]} & a^{[6]} & a^{[7]} &  a^{[10]} & a^{[11]}\end{array}\right).
\end{align*}

The semi--infinite parity--check matrix $\mathbf H$ is given by
\begin{equation*}
\mathbf H =
\left( \begin{array}{ccc}
\mathbf H_0& & \\
 \mathbf H_1& \mathbf H_0&\\
 &\mathbf H_1 &\ddots \\
&&\ddots\\
       \end{array}\right)
=\left( \begin{array}{cccccc|ccccccc}
    a^{[0]} &  a^{[1]}  & a^{[4]} &  a^{[5]} & a^{[8]} &  a^{[9]}&&&&&&&\\
    a^{[1]} &  a^{[2]}  & a^{[5]} &  a^{[6]} & a^{[9]} &  a^{[10]}&&&&&&&\\
\cline{1-12}\\[-0.9ex]
a^{[2]} &  a^{[3]}  & a^{[6]} &  a^{[7]} & a^{[10]} &  a^{[11]}&a^{[0]} &  a^{[1]}  & a^{[4]} &  a^{[5]} & a^{[8]} &  a^{[9]}&\\
a^{[3]} &  a^{[4]}  & a^{[7]} &  a^{[8]} & a^{[11]} &  a^{[12]}&a^{[1]} &  a^{[2]}  & a^{[5]} &  a^{[6]} & a^{[9]} &  a^{[10]}&\\
\cline{1-12}\\[-0.9ex]
&&&&&&a^{[2]} &  a^{[3]}  & a^{[6]} &  a^{[7]} & a^{[10]} &  a^{[11]}&\dots\\
&&&&&&a^{[3]} &  a^{[4]}  & a^{[7]} &  a^{[8]} & a^{[11]} &  a^{[12]}&\dots\\
\cline{7-12}\\[-0.9ex]
\end{array}\right).
\end{equation*}
As required by \eqref{eq:defHc}, \eqref{eq:defHr} $\mathbf H_0$, $\mathbf H_1$, $\left( \begin{array}{cc}\mathbf H_0 &\mathbf H_1\end{array}\right)$ and $\left( \begin{array}{c}\mathbf H_0 \\ \mathbf H_1\end{array}\right)$ define Gabidulin codes.

The corresponding generator matrix $\mathbf G$ consists of two submatrices $\mathbf G_0$, $\mathbf G_1$ which are $(4 \times 6)$-matrices, where the submatrices $\mathbf G_{00}$, $\mathbf G_{01}$, $\mathbf G_{10}$ are $(2 \times 6)$-matrices, since $k_1 = \nu = m_H(n-k)=2$. Hence, $\mathbf H^T$ defines a PUM code based on Gabidulin codes (to ensure that there exists such a PUM code see also Theorem~\ref{theo:HdefsPUM}). Note that the generator submatrices are not necessarily generator matrices of a Gabidulin code.
\end{example}

\subsection{Calculation of Distances}\label{subsec:calcdist}
In this section, we calculate the extended row rank distance and hence, the free rank distance $d_{free}$ and the slope $\alpha$ for the construction of Definition~\ref{def:pumgab} and dual memory $m_H=1$. We show that the free rank distance achieves the upper bound for PUM codes \eqref{eq:pumupperbound} and half the optimal slope \eqref{eq:slopeupperbound}. 

In order to estimate the extended row rank distance $d^r_{\ell}$ \eqref{eq:defextrowrank}, let us consider all paths in the set $\mathcal C^{(r)}(\ell)$. As defined in \eqref{eq:defextrowrank}, $d^r_{\ell}$ is the minimum sum rank weight \eqref{eq:defsumrankweight} of all possible words in $\mathcal C^{(r)}(\ell)$, where $\mathbf c_i \neq 0 $ for $i=0,\dots,\ell-1$ \eqref{eq:defcl}.


Let us denote $r \defeq n-k$ and the rank distances for codes defined by the following parity--check matrices:
\begin{equation*}
 d_0 \defeq d_{\rk}(\mathbf H_0), \quad d_1 \defeq d_{\rk}(\mathbf H_1), \quad d_{10}\defeq d_{\rk}\left( \begin{array}{cc}\mathbf H_1 & \mathbf H_0\end{array}\right).
\end{equation*}
Since all these matrices are Frobenius matrices \eqref{eq:linvand} with $n-k$ rows and the elements of the first row are linearly independent over $\Fq$ \eqref{eq:setH}, $d_0=d_1=d_2=d_{10}=d_{210}=n-k+1=r+1$.

\begin{theorem}
Let an $(n,k\;|\;k_1)$ (P)UM code with $m_H=1$ be given where the submatrices of $\mathbf H$ define Gabidulin codes as in Definition~\ref{def:pumgab}. 
Let the rate $R \geq 1/2$, then the extended row rank distance $d^r_{\ell}$ \eqref{eq:defextrowrank} is given by
\begin{equation}\label{eq:drowrank1}
\begin{split}
d^r_1 &= 2r+1=2(n-k)+1,\\
d^r_{\ell} &\geq \left\lceil \frac{\ell+1}{2}\right\rceil \cdot (r+1)=\left\lceil \frac{\ell+1}{2}\right\rceil \cdot (n-k+1), \quad \ell=2,3\dots.
\end{split}
\end{equation}
\end{theorem}
\begin{IEEEproof}

The analysis of the extended row rank distances for $m_H=1$ is similar to the analysis from \cite{zyablov_sidorenko_pum}.

\begin{enumerate}
\item For $\mathbf c^{(1)}=\mathbf c_0$, we have to consider the $2r\times n$ parity--check matrix
\begin{equation*}
\mathbf H^{(1)} = \left( \begin{array}{c}
\mathbf H_0\\
\mathbf H_1\\
             \end{array}\right),
\end{equation*}
i.e., $\mathbf H^{(1)}\cdot\mathbf c^{(1)T}  =\mathbf 0$. Since $\mathbf H^{(1)}$ defines a Gabidulin code with minimum rank distance $d_{\rk}=2r+1$, we have $d^r_1=2r+1$.

\item For $\mathbf c^{(2)}=\left( \begin{array}{cc}\mathbf c_{0}& \mathbf c_1\end{array}\right)$, we have to consider the $3 r \times 2n$ parity--check matrix
\begin{equation*}
\mathbf H^{(2)} = \left( \begin{array}{cc}
\mathbf H_0&\\
\mathbf H_1&\mathbf H_0\\
&\mathbf H_1\\
             \end{array}\right),
\end{equation*}
i.e., $ \mathbf H^{(2)}\cdot\mathbf c^{(2)T} =\mathbf 0$. Hence, in particular the following equations must be fulfilled:
\begin{equation*}
 \mathbf  H_0 \cdot \mathbf c_{0}^T= \mathbf 0,\qquad
\mathbf H_1 \cdot \mathbf c_{1}^T= \mathbf 0,
\end{equation*}
and hence, $d^r_2 \geq d_0+d_1 = 2(r +1)$.

\item For $\mathbf c^{(3)}=\left( \begin{array}{ccc}\mathbf c_{0}& \mathbf c_1& \mathbf c_2\end{array}\right)$, we have the same and 
$d^r_3 \geq d_0+d_1 = 2(r +1)$.

\item For $\mathbf c^{(4)}=\left( \begin{array}{cccc}\mathbf c_{0}& \mathbf c_1& \mathbf c_2&\mathbf c_3\end{array}\right)$, we have to consider the $5 r \times 4n$ parity-check matrix $\mathbf H^{(4)}$, for which 
$\mathbf H^{(4)} \cdot\mathbf c^{(4)T} =\mathbf 0$ and in particular the following holds:
\begin{equation*}
 \mathbf  H_0 \cdot \mathbf c_{0}^T= \mathbf 0,\qquad
\mathbf  H_1 \cdot \mathbf c_{1}^T+ \mathbf H_0 \cdot\mathbf c_{2}^T = \mathbf 0,\qquad
\mathbf H_1 \cdot \mathbf c_{3}^T= \mathbf 0,
\end{equation*}
and $d^r_4 \geq d_0+d_{10}+d_1 = 3(r +1)$.
\end{enumerate}
Similarly, $d^r_5 \geq d_0+d_{10}+d_1 = 3(r +1)$ and $d^r_6 \geq d_0+d_{10}+d_{10}+d_1 = 4(r +1)$. In general, we obtain \eqref{eq:drowrank1} with the same strategy.
\end{IEEEproof}

This yields the following results for the free rank distance and the slope of a PUM code based on Gabidulin codes constructed as in Definition~\ref{def:pumgab} with $m_H=1$.
\begin{theorem}\label{theo:dfreeslope}
For $R > 1/2$, the $(n,k\;|\;k_1)$ PUM code based on Gabidulin codes with $m_H= 1$ achieves the upper bound of the free rank distance $d_{free}$ and half the optimal slope $\alpha$:
\begin{align*}
d_{free} &= 2(n-k)+1=n-k+\nu+1,\\
\alpha &= \frac{n-k+1}{2}.
\end{align*}
\end{theorem}
\begin{IEEEproof}
The overall constraint length is $\nu=n-k$. Hence,
\begin{equation*}
d_{free} = \min_{\substack{\ell}} \lbrace d^r_{\ell} \rbrace = d^r_1=2(n-k)+1=n-k+\nu+1.
\end{equation*}
The slope is calculated using \eqref{eq:defslope}:
\begin{equation*}
\alpha = \lim_{\substack{\ell \rightarrow \infty}} \frac{d^r_{\ell}}{\ell}
=\frac{n-k+1}{2}.
\end{equation*}
\end{IEEEproof}


\section{Conclusion}\label{sec:concl}
We considered (P)UM codes based on Gabidulin codes. We defined general distance measures for convolutional codes based on a modified rank metric -- the sum rank metric -- and derived upper bounds for (P)UM codes based on the sum rank metric. In addition, an explicit construction of (P)UM codes based on Gabidulin codes was given and its free rank distance and slope were calculated for dual memory $m_H=1$. Our PUM construction achieves the upper bound for the free rank distance and half the optimal slope.

For future work, a decoding algorithm for these codes is of interest -- maybe similar to Dettmar and Sorger's low--complexity decoding algorithm for PUM codes based on Hamming metric block codes \cite{DettmarSorger_BMDofUM}. Also, a (P)UM construction where the \emph{generator} matrices define Gabidulin codes could be considered.

%
%

\bibliographystyle{IEEEtran}
\bibliography{antoniawachter}
\end{document}

%% file: antoniasdefs.tex
\newtheorem{theorem}{Theorem}
\newtheorem{definition}{Definition}
\newtheorem{lemma}{Lemma}
\newtheorem{corollary}{Corollary}

\newtheorem{example}{Example}

\DeclareMathOperator{\wt}{wt}
\DeclareMathOperator{\rank}{rank}
\DeclareMathOperator{\rk}{rk}

\DeclareMathOperator{\defi}{def}

\newcommand{\qed}{\hfill \mbox{\raggedright \rule{.07in}{.1in}}}

\newcommand{\defeq}{\overset{\defi}{=}}

\newcommand{\Fqs}{\mathbb F_{q^s}}
\newcommand{\Fq}{\mathbb F_{q}}
\newcommand{\F}{\mathbb F}

